\documentclass[11pt, letterpaper]{article}
\usepackage{graphicx}
\usepackage{amsfonts}
\usepackage[margin=1in]{geometry}
\usepackage{pgfplots}
\usepackage{pgfplotstable}
\usepackage{hyperref} 

\usepackage{amsmath, bm, braket}
\usepackage{amsthm}
\newtheorem{theorem}{Theorem}
\newtheorem{lemma}{Lemma}
\usepackage{algorithm}
\usepackage{setspace}
\usepackage{algorithmic}

\usepackage{caption}
\usepackage{subcaption}
\pgfplotsset{compat=1.17}
 \usepackage{url}
\usepackage{quantikz}

\usepackage[numbers,sort&compress]{natbib}

\usepackage{tikz}
\usetikzlibrary{backgrounds,fit,decorations.pathreplacing,calc}
\usetikzlibrary{circuits.ee.IEC}
\usepackage{amsmath, bm, braket}

\usepackage{bm}  
\usepackage{amsmath}       
\usepackage{amssymb}       
\usepackage{bm}   

\usepackage{setspace}
\usepackage{parskip}
\usepackage{lmodern} 

\begin{document}
\title{A Quantum Algorithm for Computing All Eigenvalues and Eigenstates of Sparse Hermitian Matrices}
\author{Krishnakanta Barik\footnote{krishnakantabarik420@gmail.com}, Goutam Paul\footnote{goutam.paul@isical.ac.in} \\  \textsuperscript{*}Cryptology and Security Research Unit, Indian Statistical Institute, Kolkata, India \\ \textsuperscript{†}Electronics and Communication Sciences Unit, Indian Statistical Institute, Kolkata, India}

\date{\today}

\maketitle

\begin{abstract}
Computation of eigenvalues and eigenvectors of a sparse Hermitian matrix is a fundamental problem in algebra, arising both as a problem of independent interest and as a key subroutine in a wide range of computational and scientific applications.
In this work, we develop a quantum algorithm that determines the spectrum of a sparse Hermitian matrix, defined as the set of its distinct eigenvalues, and for each eigenvalue, prepares a corresponding eigenvector as a quantum state in its eigenspace.
The eigenvalues are represented as binary strings with a prescribed precision, while the corresponding eigenvectors are kept as quantum states, eliminating the need to reconstruct their classical counterparts.
The algorithm instead enables estimation of expectation values $\braket{x|M|x}$, where $\ket{x}$ is a prepared eigenstate and $M$ is an observable.
For an $n \times n$ $s$-sparse Hermitian matrix $H$, dense methods based on Householder tridiagonalization and QR iteration compute all eigenvalues and eigenvectors in $\mathcal{O}(n^3)$ time.
Exploiting sparsity, the Lanczos method computes all eigenvalues in $\mathcal{O}(n^2 s)$ time in exact arithmetic, but recovering all eigenvectors still requires $\mathcal{O}(n^3)$ time.
In contrast, our algorithm runs in time $\tilde{\mathcal{O}}(n s^2 \log^2 n / \Delta_{\min})$, where $\Delta_{\min}$ denotes the minimum spectral gap of the normalized matrix.
For inverse gaps up to $\Delta_{\min}^{-1} = \mathcal{O}(n)$, the most favorable scaling for $n$ distinct eigenvalues, attained by nearly uniformly spaced spectra such as those of perfect-state-transfer spin chains and the Sylvester-Kac matrix, and for sparsity up to $s = o(\sqrt{n}/\log n)$, the proposed algorithm achieves an improved asymptotic runtime over the classical $\mathcal{O}(n^3)$ approach for computing all eigenvalues and eigenvectors, under standard sparse-access assumptions.
\end{abstract}

\maketitle

\section{Introduction}
This spectral information associated with a matrix~\cite{Horn_Johnson_2012} plays an important role in various areas of computer science, including graph analysis~\cite{chung1997spectral} and combinatorial optimization~\cite{10.1145/1536414.1536452}.
A well-known classical method for computing the spectrum of a dense matrix is the tridiagonal QR algorithm~\cite{golub2013matrix, trefethen1997numerical}.
It first reduces a matrix to tridiagonal form using Householder reflections, requiring approximately $\frac{4}{3}n^3$ floating-point operations (flops) for a real symmetric matrix.
Once the matrix is reduced to tridiagonal form, all of its eigenvalues can be computed using the implicitly shifted QR algorithm in $\mathcal{O}(n^2)$ operations. 
Therefore, the overall computational complexity remains $\mathcal{O}(n^3)$, dominated by the tridiagonalization step.
For an $n \times n$ $s$-sparse Hermitian matrix, the Lanczos algorithm~\cite{lanczos1950iteration} computes all eigenvalues in $\mathcal{O}(n^2 s)$ operations in exact arithmetic: $n$ iterations of $\mathcal{O}(ns)$ sparse matrix-vector products, followed by an $\mathcal{O}(n^2)$ tridiagonal eigensolve~\cite{golub2013matrix}. 
Recovering all $n$ eigenvectors additionally requires storing the Lanczos basis $Q$ and forming $X = QY$, where $Y$ holds the tridiagonal eigenvectors, obtainable in $\mathcal{O}(n^2)$ time via MRRR~\cite{dhillon2006design}. 
This dense product costs $\mathcal{O}(n^3)$ operations, or $\mathcal{O}(n^{\omega})$ with fast matrix multiplication~\cite{alman2025more}, where $\omega < 2.372$.
In floating point, the Lanczos vectors lose orthogonality as Ritz values converge~\cite{paige1980accuracy}, so full reorthogonalization, at an additional $\mathcal{O}(n^3)$ cost, is needed for an accurate, orthogonal set of eigenvectors~\cite{parlett1998symmetric,saad2011numerical}.

Two major components of quantum computing~\cite{nielsen2010quantum, mermin2007quantum, horowitz2019quantum} are qubits and quantum gates that are analogous to classical bits and classical gates, respectively.
Quantum computing has the potential to provide advantages~\cite{grover1996fast, shor1994algorithms} for certain classical computational tasks by exploiting fundamental quantum-mechanical properties, such as superposition and entanglement.
Superposition allows a quantum state to be represented as a linear combination of tensor products of qubits, while entanglement enables correlations between qubits that have no direct classical counterpart.
These properties enable the development of quantum algorithms that can address certain computational problems more efficiently than their classical counterparts, particularly in areas such as optimization, simulation, and numerical linear algebra.
Current quantum computing platforms are characterized by the noisy intermediate-scale quantum (NISQ) era~\cite{Preskill2018quantumcomputingin}, featuring increasingly capable quantum devices with more qubits and improved computational reliability.

In this work, we present a quantum algorithm for determining the spectrum of a sparse Hermitian matrix $H$, together with the corresponding eigenvectors.
We focus specifically on reconstructing the distinct spectrum of the Hamiltonian, defined as the set of unique eigenvalues $\operatorname{spec}(H) = \{\lambda_1, \dots, \lambda_d\}$, where $d \le n$. 
When an eigenvalue $\lambda_j$ is degenerate (appearing with multiplicity $m_j > 1$), measuring an estimated eigenvalue within its tolerance bin collapses the state register into the corresponding eigenvector space $V_{\lambda_j} = \operatorname{span}\{\lvert x\rangle \mid H\lvert x\rangle = \lambda_j \lvert x\rangle\}$.
The eigenvalues are obtained in classical binary-string form with a prescribed precision, providing an explicit classical representation of the spectrum. In contrast, the corresponding eigenvectors are maintained in quantum-state form, rather than being explicitly reconstructed as classical vectors.
Instead, relevant properties of the corresponding eigenvectors can be characterized by directly estimating expectation values of observables of the form $\braket{x|M|x}$, where $\ket{x}$ denotes the quantum state corresponding to an eigenvector and $M$ is a given observable.
Our algorithm is based on the Quantum Phase Estimation (QPE) algorithm and consists of three main stages.
First, the given sparse Hermitian matrix is mapped to a corresponding unitary operator suitable for phase estimation.
Second, QPE is applied to this unitary operator multiple times, with a predefined computational basis state used as the input state of the eigenvector register in each execution. 
Finally, in each execution, the eigenvalue register is measured to obtain the corresponding eigenvalue in classical binary-string form. At the same time, the associated eigenvector is retained as a quantum state in the eigenvector register.
The overall runtime of the proposed algorithm is $\tilde{\mathcal{O}}\left( \frac{ns^2\log^2(n)}{\Delta_{\min}}\right)$, where $\Delta_{\min}$ denotes the minimum spectral gap.
For inverse minimum spectral gaps up to $\Delta_{\min}^{-1} = \mathcal{O}(n)$, and under the stated sparsity and matrix-access assumptions, this complexity yields an improved asymptotic scaling compared with the $\mathcal{O}(n^3)$ runtime of conventional classical methods for computing eigenvalues and eigenvectors.
Such spectra arise, for example, in spin chains engineered for perfect quantum state transfer, whose single-excitation Hamiltonians have equally spaced eigenvalues~\cite{christandl2004perfect}, in spin-$j$ angular momentum operators and the closely related Sylvester-Kac matrix~\cite{kac1947random,clement1959class}, and in weakly perturbed diagonal Hamiltonians with well-separated energy levels~\cite{Horn_Johnson_2012}.

The remainder of this paper is organized as follows. 
In Section~\ref{Related Work}, we review existing quantum algorithms for computing the eigenvalues and eigenvectors of matrices. 
Section~\ref{Quantum Algorithm for   Spectrum Estimation} formally presents the proposed quantum algorithm for spectrum estimation.
Finally, Section~\ref{Conclusion and Future Directions} concludes the paper and discusses potential directions for future research.

\section{Related Work}
\label{Related Work}
Three major quantum-computing approaches have been employed to address eigenvalue estimation: adiabatic quantum computation, variational quantum algorithms, and quantum phase estimation. 
In the following, we discuss these approaches, highlighting their respective advantages and challenges.

\subsection{Quantum Adiabatic Computing}
\label{Quantum Adiabatic Computing}
Quantum adiabatic computation is a well-established approach for estimating eigenvalues of a sparse Hermitian matrix, based on the quantum adiabatic theorem.
Let us briefly review the quantum adiabatic theorem~\cite{pastorello2022concise, RevModPhys.90.015002, farhi2000quantumcomputationadiabaticevolution, PhysRevA.89.022342}.
Assume that $H(t)$ is a time-dependent Hamiltonian of a given quantum system, defined over the interval $0 \leq t \leq T$, where $H(0)$ and $H(T)$ denote the initial and final Hamiltonians, respectively.
If the system is initially prepared in the ground state of $H(0)$ and $H(t)$ evolves sufficiently slowly, the system will remain in the instantaneous ground state of $H(t)$. 
The rate of evolution of $H(t)$ must be sufficiently slow relative to the energy gap between the ground state and the first excited state.
We now illustrate how the quantum adiabatic theorem can be used to compute the ground-state energy, which corresponds to the minimum eigenvalue of the Hamiltonian.
The procedure consists of the following three main steps~\cite{pastorello2022concise}.

\begin{itemize}
    \item A quantum system is prepared in the known ground state of an initial Hamiltonian $H(0)$. The final Hamiltonian $H(T)$, or problem Hamiltonian, is constructed such that its ground state encodes the solution to the given problem.
    \item A time-dependent Hamiltonian interpolating between $H(0)$ and $H(T)$ is defined as~\cite{PhysRevA.89.022342}
    \begin{equation}
    H(t)=\left(1-m(t)\right)H(0)+m(t)H(T), \qquad 0\leq t\leq T,
    \nonumber
    \end{equation}
    where $m:[0,T]\rightarrow[0,1]$ is a smooth, monotonic function satisfying $m(0)=0$ and $m(T)=1$.
    \item Finally, the quantum system is measured after the adiabatic evolution. The measurement yields the ground state of $H(T)$, which encodes the solution to the given problem.
\end{itemize}
The correctness of this algorithm follows from the quantum adiabatic theorem.
A major limitation of adiabatic state preparation is that the minimum spectral gap~\cite{PhysRevA.89.022342} can become extremely small as the system size increases, particularly when the initial and final Hamiltonians have substantially different eigenstate structures or when the evolution encounters quantum phase transitions. 
In such cases, maintaining adiabaticity may require a very long evolution time, making the approach computationally demanding for large systems.

\subsection{Variational Quantum Eigensolver}

The Variational Quantum Eigensolver (VQE)~\cite{peruzzo2014variational, RevModPhys.94.015004} is a Variational Quantum Algorithm (VQA)~\cite{cerezo2021variational} widely used for estimating the eigenvalues and corresponding eigenvectors of a sparse Hermitian matrix.
It is a hybrid quantum-classical algorithm that combines a parameterized quantum circuit with classical optimization, where the quantum circuit prepares a trial state and estimates its energy. In contrast, the classical optimizer updates the circuit parameters to minimize the energy.
The procedure consists of the following four main components.
\begin{itemize}
    \item An operator: typically a Hamiltonian $H$ that describes a property of the system whose lowest (minimum) eigenvalue we wish to find.
    \item An ansatz: a quantum circuit that prepares a quantum state approximating the eigenvector of interest.
    \item An estimator: a method for estimating the expectation value of the operator $H$ with respect to the current variational quantum state.
    \item A classical optimizer: an algorithm that updates the variational parameters to minimize the cost function.
\end{itemize}
Although VQE provides an effective framework for estimating eigenvalues and preparing the corresponding eigenstates, its primary objective is typically to determine the ground-state energy (the minimum eigenvalue) and the corresponding eigenstate.
The performance of VQE depends on the choice of ansatz, initialization, and classical optimization procedure, and it may be affected by barren plateaus and local minima~\cite{cerezo2021variational}. 
Consequently, VQE does not provide a systematic procedure for obtaining the complete spectrum and all corresponding eigenstates of a matrix. In contrast, our work focuses specifically on estimating the complete spectrum and the corresponding eigenstates.

\subsection{Quantum Phase Estimation Algorithm}
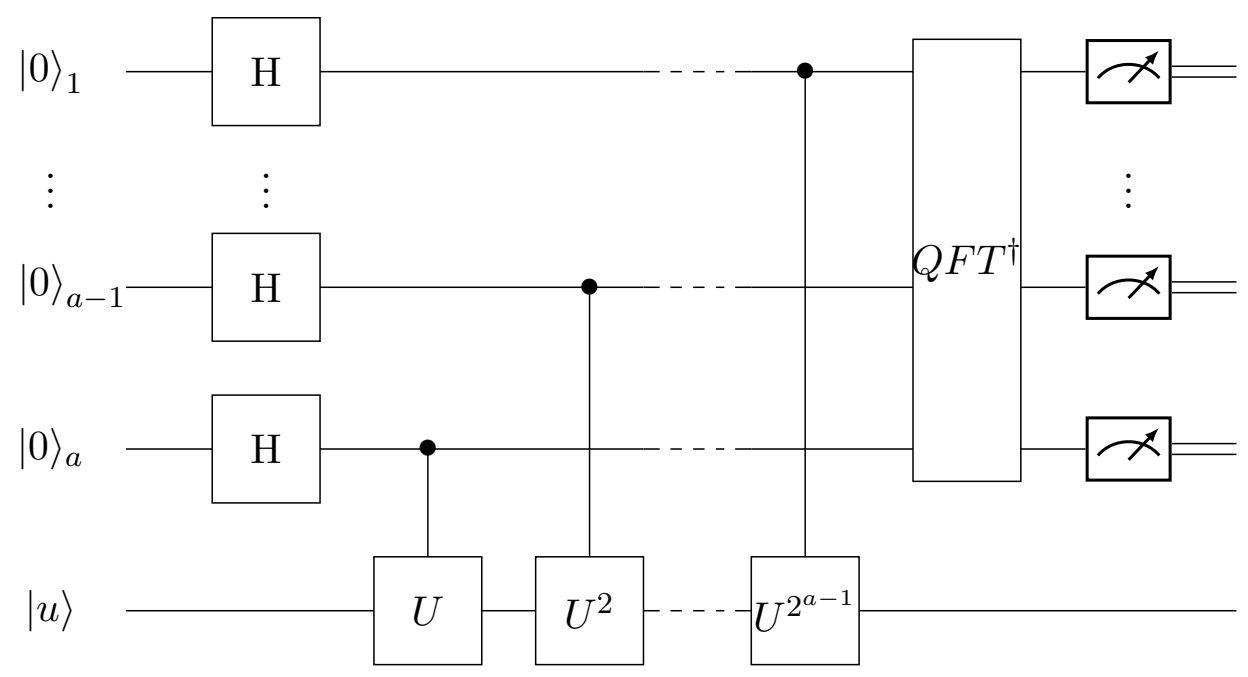
\begin{figure}[]
  \centering
  \resizebox{0.5\linewidth}{!}{%
    \begin{tikzpicture}
\tikzstyle{surround} = [fill=lime!10,thick,draw = black,rounded corners = 2mm]
\draw node at (0,0) (q1) {$\ket{0}_{1}$};
\draw (0.7,0)--(1.5,0);
\draw (1.5,0.5) rectangle (2.5,-0.5);
\draw node at (2,0) {H};
\draw (2.5,0)--(5.5,0);
\draw[dashed] (5.75,-0) -- (6.25,-0);
\draw (5.5,-0) -- (5.65,-0);
\draw (6.35,-0) -- (6.5,-0);
\draw (6.5,0)--(8,0);

\draw node at (0,-1) {$\vdots$};
\draw node at (2,-1) {$\vdots$};
\draw node at (0.2,-2) (q2) {$\ket{0}_{a-1}$};
\draw (0.7,-2)--(1.5,-2);
\draw (1.5,-1.5) rectangle (2.5,-2.5);
\draw node at (2,-2) {H};
\draw (2.5,-2)--(5.5,-2);
\draw[dashed] (5.75,-2) -- (6.25,-2);
\draw (5.5,-2) -- (5.65,-2);
\draw (6.35,-2) -- (6.5,-2);
\draw (6.5,-2)--(8,-2);

\draw node at (0,-3.5) (q3) {$|0\rangle_{a}$};
\draw (0.7,-3.5)--(1.5,-3.5);
\draw (1.5,-3) rectangle (2.5,-4);
\draw node at (2,-3.5) {H};
\draw (2.5,-3.5)--(5.5,-3.5);
\draw[dashed] (5.75,-3.5) -- (6.25,-3.5);
\draw (5.5,-3.5) -- (5.65,-3.5);
\draw (6.35,-3.5) -- (6.5,-3.5);
\draw (6.5,-3.5)--(8,-3.5);
\draw node at (0,-5) (qu) {$\ket{u}$};
\draw (0.7,-5)--(3,-5);
\draw (3,-4.5) rectangle (4,-5.5);
\draw node at (3.5,-5) {$U$};
\draw (3.5,-3.5) -- (3.5,-4.5);
\draw (4.5,-4.5) rectangle (5.5,-5.5);
\draw (4,-5) -- (4.5,-5);
\draw node at (5,-5) {$U^2$};
\draw node at (5,-2) {$\bullet$};
\draw (5,-2) -- (5,-4.5);
\draw node at (3.5,-3.5) {$\bullet$};
\draw[dashed] (5.75,-5) -- (6.25,-5);
\draw (5.5,-5) -- (5.65,-5);
\draw (6.35,-5) -- (6.5,-5);
\draw (6.5,-4.5) rectangle (7.5,-5.5);
\draw node at (7,-5) {$U^{2^{a-1}}$};
\draw node at (7,0) {$\bullet$};
\draw (7,0) -- (7,-4.5);
\draw (8,0.3) rectangle (9,-3.8);
\draw node at (8.5,-1.75) {$QFT^\dagger$};
\draw (9,0) -- (9.6,0);
\draw (9,-2) -- (9.6,-2);
\draw (9,-3.5) -- (9.6,-3.5);
\node[meter] at (10,0) {};
\node[meter] at (10,-2) {};
\node[meter] at (10,-3.5) {};
\draw node at (10,-1) {$\vdots$};
\draw (10.4,0.05) -- (11,0.05);
\draw (10.4,-0.05) -- (11,-0.05);
\draw (10.4,-1.95) -- (11,-1.95);
\draw (10.4,-2.05) -- (11,-2.05);
\draw (10.4,-3.45) -- (11,-3.45);
\draw (10.4,-3.55) -- (11,-3.55);
\draw (7.5,-5) -- (11,-5);
\end{tikzpicture}%
  }
  \caption{Circuit diagram of the QPE algorithm. The circuit takes the eigenstate $\ket{u}$ of a unitary operator $U$ and $a$ ancillary qubits to achieve the desired eigenvalue precision. Hadamard gates are first applied to the ancillary qubits, followed by controlled applications of $U$ with increasing powers. The inverse Quantum Fourier Transform (QFT) is then applied to the ancillary qubits, which are subsequently measured to obtain an estimate of the eigenvalue.}
  \label{fig:qpe_circuit}
\end{figure}

Quantum Phase Estimation (QPE)~\cite{J__2022, Nielsen_Chuang_2010} is a quantum algorithm for estimating the eigenvalue (eigenphase) of a given unitary operator.
Let $U$ be a unitary operator and $\ket{u}$ an eigenvector of $U$. The Quantum Phase Estimation algorithm estimates the eigenvalue corresponding to the eigenstate $\ket{u}$.
Since $U$ is unitary, all its eigenvalues have unit modulus. Thus, 
\begin{equation*}
    U\ket{u}=e^{2\pi i\lambda}\ket{u}, \hspace{1cm} \text{for some } \lambda\in[0, 1).
\end{equation*}
In general, the algorithm estimates an approximate value of $\lambda$, referred to as the eigenphase.
In the circuit diagram~\ref{fig:qpe_circuit}, the first $a$ qubits determine the precision with which the eigenvalue can be represented; thus, $a$ depends on the desired eigenvalue precision~\cite{Zaman_2023}.
To obtain $\lambda$ accurate to $b$ bits with a success probability of at least $1-\epsilon$, one can choose $a$ such that
\begin{equation*}
    a=b+\left\lceil\log\left(2+\frac{1}{2\epsilon}\right)\right\rceil.
\end{equation*}

The initial state of the system is $\ket{0}^{\otimes a}\otimes\ket{u}$. Applying the Hadamard operation $H^{\otimes a}$ to the first $a$ qubits transforms the state into 
\begin{equation*}
    \frac{1}{\sqrt{2^{a}}}\sum_{j=0}^{2^a-1}\ket{j}\otimes\ket{u}.
\end{equation*}
Next, controlled-$U$ operations with increasing powers $2^0, 2^1, 2^2,\ldots,2^{a-1}$ are applied to the state qubits $\ket{u}$, controlled by the first $a$ qubits. The resulting state is
\begin{equation*}
    \frac{1}{\sqrt{2^{a}}}\sum_{k=0}^{2^a-1}e^{2\pi i\lambda k}\ket{k}\otimes\ket{u}, \quad \text{since }U^k\ket{u}=e^{2\pi i\lambda k}\ket{u}.
\end{equation*}
Finally, applying the inverse Quantum Fourier Transform (QFT) to the first $a$ qubits yields the state
\begin{equation*}
    \frac{1}{2^a}\sum_{k=0}^{2^a-1}\sum_{z=0}^{2^a-1}
e^{2\pi i k\left(\lambda-\frac{z}{2^a}\right)}
\ket{z}\otimes\ket{u}.
\end{equation*}
Upon measuring the first $a$ qubits, suppose the measurement outcome is the binary string $\psi$. Then $\psi$ provides an approximation to $2^a\lambda$, giving $\lambda\approx\frac{\psi}{2^a}$. Thus, we obtain an estimate of the eigenvalue corresponding to the eigenvector $\ket{u}$ of the unitary operator $U$.

A key limitation of the standard QPE framework is that it assumes the input state is an eigenstate $\ket{u}$ of the unitary operator $U$, satisfying $U\ket{u}=e^{2\pi i\lambda}\ket{u}$.
Consequently, QPE estimates the eigenphase corresponding to a given eigenstate but does not, by itself, provide a procedure for identifying the set of eigenvalues when the eigenvectors are not known in advance. 
If an arbitrary input state $\ket{\phi}=\sum_i c_i\ket{u_i}$ is used, where $\ket{u_i}$ denotes the eigenvectors of $U$, QPE returns the eigenphase $\lambda_i$ corresponding to $\ket{u_i}$ with probability $|c_i|^2$. Hence, an eigenvalue whose corresponding eigenvector has zero overlap with $\ket{\phi}$ cannot be observed.
This motivates our approach of using a predefined set of computational basis states as input states across multiple QPE executions.
Since every eigenvector has a nonzero overlap with at least one predefined computational basis state, this strategy provides a systematic way to access all eigenvalue-eigenvector pairs without requiring prior knowledge of the eigenvectors.

The idea of applying phase estimation to an input state that is not an eigenstate, so that eigenvalues are sampled according to the overlap of the input with each eigenvector, goes back to Abrams and Lloyd~\cite{abrams1999quantum}. 
Our work builds on this idea in three ways:
(i)~we use a complete orthonormal basis of input states and derive an explicit bound on the number of repetitions needed to recover every distinct eigenvalue with probability at least $1-\delta$ (Theorem~\ref{theorem_4}); 
(ii)we give a
normalization and phase-recovery procedure that determines the sign of each eigenvalue (Section~\ref{Eigenvalue Scaling and Phase Recovery}); and 
(iii)~we analyze the overall complexity and compare it with classical eigensolvers for sparse matrices (Section~\ref{Complexity Analysis}).
We are now ready to present our algorithm for determining the spectrum and the corresponding eigenvectors.

\section{Quantum Algorithm for Spectrum Estimation}
\label{Quantum Algorithm for Spectrum Estimation}

Let $H\in\mathbb{C}^{n\times n}$ be a sparse Hermitian matrix. We develop a quantum algorithm to determine the eigenvalues of $H$ and obtain their corresponding eigenvectors as quantum states.
We first normalize the given matrix $H$ using its induced $1$-norm such that all eigenvalues of the resulting matrix lie strictly within the interval $(-0.5, 0.5)$.
Now, for the normalized sparse Hermitian matrix, we define the corresponding unitary operator as
\begin{equation*}
    U_H=e^{2\pi iH}.
\end{equation*}
We denote the matrix and its normalized form by the same notation, $H$.
We assume that $H$ is accessed through standard sparse-access oracles~\cite{Berry_2006}: one oracle that, given a row index $j$ and an integer $\ell \le s$, returns the column index of the $\ell$-th nonzero entry in row $j$, and another that, given a pair $(j,k)$, returns the entry $h_{jk}$.
Then we apply the Quantum Phase Estimation (QPE) algorithm to this unitary $U_H$.
We employ $a$ qubits in the phase-estimation register to represent each eigenphase, where the desired estimation precision and success probability determine $a$.
The relationship between $a$, the required number of accuracy bits, and the success probability is established later.

Rather than assuming prior knowledge of the eigenvectors, we execute QPE multiple times, using the $n$ states of a complete orthonormal basis
\begin{equation*}
    \{\ket{v_i}: i=1,2,\ldots,n\},
\end{equation*}
as input states for the eigenvector register.
This choice enables the QPE procedure to access the set of eigenvalue-eigenvector pairs, as described in the subsequent analysis.
The proposed quantum algorithm proceeds through the following sequence of steps. The overall procedure consists of constructing the unitary operator corresponding to the normalized sparse Hermitian matrix $H$, preparing the eigenvector register in a set of predefined basis states, applying QPE to estimate the corresponding eigenvalue, and measuring the phase register to obtain the eigenvalues in classical form.
The eigenvector register retains the corresponding eigenstates as quantum states. 
The detailed steps of the Algorithm~\ref{algorithm} are presented below.

\begin{algorithm}[]
\caption{Quantum algorithm for finding all eigenvalues and their corresponding eigenvectors.}
\label{algorithm}
\ \\
\textbf{Input:} An $n\times n$ Hermitian $s$-sparse matrix $H$, an integer $b$ specifying the number of bits used for eigenvalue approximation, and two parameters $\epsilon$ and $\delta$ specifying the desired failure probability.\\
\textbf{Output:} $b$-bit approximations $\tilde{\lambda}_1, \dots,
\tilde{\lambda}_d$ of all distinct eigenvalues of $H$, and, for each
$\tilde{\lambda}_j$, a quantum state in the eigenspace of $\lambda_j$.

\begin{algorithmic}[1]
\setstretch{1.5}

\STATE  \textbf{Normalize the Hamiltonian:} Compute the induced $1$-norm $\|H\|_1$. Set a scaling factor $\alpha = 2 \|H\|_1 + \eta$ (for an arbitrary small constant $\eta > 0$), and construct the normalized Hamiltonian such that all its eigenvalues strictly lie within the interval $(-0.5, 0.5)$.

\STATE \textbf{Hamiltonian Simulation:} Simulate the unitary operator $U_H=e^{2\pi iH}.$

\STATE \textbf{Construct Basis States:} Construct an orthonormal basis $\{\ket{v_i}: i\in\{1,2,\ldots,n\}\}.$

\STATE \textbf{Allocate Approximation Qubits:} To obtain a $b$-bit approximation of the eigenvalues with success probability at least $1-\epsilon$, allocate $a=b+\left\lceil\log\left(2+\frac{1}{2\epsilon}\right)\right\rceil$ ancillary qubits and initialize them in the state $\ket{0}^{\otimes a}$.

\STATE \textbf{Repeated Quantum Phase Estimation:}
Set $R=\left\lceil\frac{\ln(\frac{n}{\delta})}{1-\epsilon}\right\rceil,$ where $\delta\in(0,1)$ is the desired upper bound on the probability of failing to recover at least one distinct eigenphase.
For each computational basis state $\ket{v_i}$, $i=1,\ldots,n$, execute QPE independently $R$ times, using $a$ approximation qubits initialized to $\ket{0}^{\otimes a}$ and $\ket{v_i}$ as the initial state of the eigenvector register. Record the measured phase estimate from every execution.

\STATE \textbf{Measurement and Eigenvalue Recovery:}
For each QPE execution, measure the $a$ approximation qubits to obtain a phase estimate $\tilde{\lambda}$. Convert each measured phase to the corresponding normalized eigenvalue, correcting for phase wrapping when necessary. Rescale the estimate according to the normalization performed in Step 1 to obtain an approximation of the original eigenvalue. Retain the corresponding post-measurement state of the eigenvector register.
\end{algorithmic}
\end{algorithm} 
In the following section, we establish the correctness of this algorithm.

\subsection{Spectral Correspondence Between $H$ and $U_H$}
We now establish the relationship between the eigenvalues and eigenvectors of the sparse Hermitian matrix $H$ and those of the corresponding unitary operator $U_H$.
Let the given sparse Hermitian matrix $H$ have the spectral decomposition
\begin{equation*}
H=\sum_{j=1}^{n}\lambda_j\ket{u_j}\bra{u_j},
\end{equation*}
where $\lambda_j$ and $\ket{u_j}$ denote the eigenvalues and the corresponding orthonormal eigenvectors of $H$, respectively.
The unitary operator associated with $H$ is defined as
\begin{equation*}
    U_H=e^{2\pi iH}
=\sum_{j=1}^{n}e^{2\pi i\lambda_j}
\ket{u_j}\bra{u_j}.
\end{equation*}
Consequently,
\begin{equation*}
    U_H\ket{u_j}
=e^{2\pi i\lambda_j}\ket{u_j}.
\end{equation*}
Thus, $H$ and $U_H$ have the same eigenvectors, while each eigenvalue $\lambda_j$ of $H$ is mapped to the corresponding eigenvalue $e^{2\pi i\lambda_j}$ of $U_H$.
Conversely, if the eigenvalues of $H$ are restricted to an interval over which the mapping
\begin{equation}
    \lambda \mapsto e^{2\pi i\lambda}
    \label{Eigen_value}
\end{equation}
is injective, then the eigenvalues of $H$ can be uniquely recovered from those of $U_H$.

\subsection{Eigenvalue Scaling and Phase Recovery}
\label{Eigenvalue Scaling and Phase Recovery}
We now derive the conditions under which the mapping in Equation~\ref{Eigen_value} is injective, ensuring that the eigenvalues of $H$ can be uniquely recovered from those of $U_H$.
QPE estimates the phase $\psi$ of a unitary eigenvalue in the interval $\psi\in[0,1)$.
In our construction, the unitary operator is defined as
\begin{equation*}
    U_H=e^{2\pi iH},
\end{equation*}
where the eigenvalues of the sparse Hermitian matrix $H$ are real-valued and may be positive or negative, and need not lie within the interval $[0,1)$.
Since
\begin{equation*}
    U_H\ket{u_j}=e^{2\pi i\lambda_j}\ket{u_j},
\end{equation*}
QPE estimates $\lambda_j$ only the interval $[0,1)$.
Consequently, different eigenvalues that differ by an integer, such as $\lambda_j=0.3$ and $\lambda_j=-0.7$, produce the same phase and cannot be distinguished directly.
Therefore, an appropriate scaling or shifting of the sparse Hermitian matrix is required to ensure that the eigenvalues can be uniquely recovered from the estimated phases.
For this reason, we scale the sparse Hermitian matrix so that all its eigenvalues lie within the interval $(-0.5,0.5)$. 
After applying QPE to the resulting unitary operator, positive eigenvalues in the interval $[0,0.5)$ are directly represented by phases in $[0,0.5)$, while negative eigenvalues in $(-0.5,0)$ are mapped to phases in $(0.5,1)$. 
Thus, the sign of each eigenvalue can be determined from the estimated phase, allowing the original eigenvalues to be recovered by appropriately interpreting the QPE output.
To scale the Hamiltonian, we use its induced $1$-norm (maximum absolute column sum), defined as
\begin{equation*}
    \|H\|_1 = \max_{1 \le j \le n} \sum_{i=1}^n |h_{ij}|.
\end{equation*}
This norm provides a quick, easy upper bound on the largest eigenvalue, ensuring that all eigenvalues fit within the target interval. The following lemma shows how quickly this norm can be computed and confirms the resulting bounds on the eigenvalues.

\begin{lemma}
Let $H \in \mathbb{C}^{n \times n}$ be an $s$-sparse Hermitian matrix. The induced $1$-norm $\|H\|_1 = \max_{j} \sum_{i=1}^n |h_{ij}|$ can be computed in $\mathcal{O}(ns)$ time. Setting 
\[
\alpha = 2 \|H\|_1 + \eta
\]
for any fixed constant $\eta > 0$, the scaled Hamiltonian $\tilde{H} = H / \alpha$ satisfies
\[
\mathrm{spec}(\tilde{H}) \subset (-0.5, 0.5).
\]
\end{lemma}

\begin{proof}
Since $H$ is Hermitian, its induced $1$-norm and $\infty$-norm coincide $\|H\|_\infty = \|H\|_1$. By the standard norm inequality $\|H\|_2 \le \sqrt{\|H\|_1 \|H\|_\infty}$, we have $\|H\|_2 \le \|H\|_1$. 
Consequently, for any eigenvalue $\lambda \in \mathrm{spec}(H)$,
\[
|\lambda| \le \|H\|_2 \le \|H\|_1.
\]
For any eigenvalue $\tilde{\lambda} \in \mathrm{spec}(\tilde{H})$, it follows that
\[
|\tilde{\lambda}| = \frac{|\lambda|}{\alpha} \le \frac{\|H\|_1}{2\|H\|_1 + \eta} < 0.5,
\]
which guarantees that all eigenvalues of $\tilde{H}$ lie strictly within $(-0.5, 0.5)$. Finally, since each column of $H$ has at most $s$ non-zero elements, computing $\|H\|_1$ requires evaluating $n$ column sums of size at most $s$, taking $\mathcal{O}(ns)$ classical operations.
\end{proof}

Standard Quantum Phase Estimation restricts its measured phase to the principal interval $[0, 1)$, meaning negative eigenvalues cannot be read off directly without ambiguity. By mapping the spectrum of the scaled Hamiltonian strictly inside $(-0.5, 0.5)$, negative eigenvalues wrap predictably into the upper half of the unit interval, as formalized below.

\begin{theorem}
Let $\lambda\in(-0.5,0)$ be an eigenvalue of the scaled sparse Hermitian matrix $H$, with corresponding eigenvector $\ket{u}$. Under the unitary operator $U_H$, the phase $\lambda$ is represented by the equivalent phase $1+\lambda\in(0.5,1)$. Consequently, if QPE with $a$ precision qubits produces the measurement outcome $\psi$, then
\begin{equation*}
\frac{\psi}{2^a}\approx 1+\lambda,
\end{equation*}
and the original negative eigenvalue can be recovered as
\begin{equation*}
\lambda\approx \frac{\psi}{2^a}-1.
\end{equation*}
Thus, after scaling the eigenvalues of $H$ into $(-0.5,0.5)$, negative eigenvalues are uniquely identifiable from the QPE output as those whose estimated phases lie in $(0.5,1)$.
\end{theorem}

\begin{proof}
Consider an eigenvalue $\lambda\in(-0.5,0)$ with corresponding eigenvector $\ket{u}$. Immediately before measurement, the QPE circuit produces the state
\begin{equation*}
\frac{1}{2^a}\sum_{k=0}^{2^a-1}\sum_{z=0}^{2^a-1}
e^{2\pi i k\left(\lambda-\frac{z}{2^a}\right)}
\ket{z}\otimes\ket{u}.
\end{equation*}
Since QPE represents phases in the interval $[0,1)$, we express the negative phase $\lambda$ as the equivalent phase $1+\lambda$. For every integer $k$,
\begin{equation*}
e^{2\pi i k\lambda}
=
e^{2\pi i k(1+\lambda)},
\end{equation*}
because $e^{2\pi i k}=1$. Therefore, the QPE state can equivalently be written as
\begin{equation*}
\begin{aligned}
\frac{1}{2^a}\sum_{k=0}^{2^a-1}\sum_{z=0}^{2^a-1}
e^{2\pi i k\left(\lambda-\frac{z}{2^a}\right)}
\ket{z}\otimes\ket{u}
&=
\frac{1}{2^a}\sum_{k=0}^{2^a-1}\sum_{z=0}^{2^a-1}
e^{2\pi i k\left(1+\lambda-\frac{z}{2^a}\right)}
\ket{z}\otimes\ket{u}.
\end{aligned}
\end{equation*}
Because $\lambda\in(-0.5,0)$, the equivalent phase satisfies $1+\lambda\in(0.5,1)$.
By the standard QPE analysis, measurement of the first $a$ qubits yields an integer $\psi$ such that
\begin{equation*}
\frac{\psi}{2^a}\approx 1+\lambda.
\end{equation*}
Hence, when the measured phase lies in $(0.5,1)$, it represents a negative eigenvalue. The original eigenvalue is therefore recovered by subtracting $1$ from the measured phase
\begin{equation*}
\lambda\approx\frac{\psi}{2^a}-1.
\end{equation*}
Thus, scaling the eigenvalues of $H$ into the interval $(-0.5,0.5)$ separates positive and negative eigenvalues in the QPE output: positive eigenvalues are represented directly in $[0,0.5)$, whereas negative eigenvalues are represented by phases in $(0.5,1)$ and can be recovered by subtracting $1$ from the measured phase.
\end{proof}

Having established how individual eigenvalues are mapped under phase estimation, we now analyze the routine's action on an arbitrary input superposition. The following theorem bounds the probability of obtaining an accurate eigenvalue estimate alongside its corresponding eigenstate in the presence of phase truncation errors.

\begin{theorem}\label{thm:qpe-superposition}
Let the input state be
$|0\rangle^{\otimes a} \otimes \sum_{j=1}^{n} c_j |u_j\rangle$, where
$\sum_{j=1}^{n} |c_j|^2 = 1$. After applying quantum phase estimation (QPE),
the resulting state is, up to the phase-estimation error,
\begin{equation*}
    \sum_{j=1}^{n} c_j\, |\tilde{\lambda}_j\rangle |u_j\rangle,
\end{equation*}
where $|\tilde{\lambda}_j\rangle$ denotes the phase-register state encoding an estimate of the eigenphase $\lambda_j$ associated with $|u_j\rangle$. If the number of phase-estimation qubits is chosen as $a = b + \left\lceil \log_2\!\left(2 + \frac{1}{2\epsilon}\right) \right\rceil$, then, for each $j$, the probability of obtaining the eigenstate component $|u_j\rangle$ together with an estimate of $\lambda_j$ accurate to $b$ bits is at least $|c_j|^2(1-\epsilon)$.
\end{theorem}
 
\begin{proof}
The initial state is
$|0\rangle^{\otimes a} \otimes \sum_{j=1}^{n} c_j |u_j\rangle$, where
$U|u_j\rangle = e^{2\pi i \lambda_j}|u_j\rangle$ and $\lambda_j \in [0,1)$
denotes the corresponding eigenphase. Applying Hadamard gates to the phase-estimation register gives
\begin{equation*}
    \frac{1}{\sqrt{2^a}} \sum_{j=1}^{n} \sum_{k=0}^{2^a-1} c_j\, |k\rangle |u_j\rangle .
\end{equation*}
Since $U^k|u_j\rangle = e^{2\pi i k \lambda_j}|u_j\rangle$, the controlled
powers of $U$ produce
\begin{equation*}
    \frac{1}{\sqrt{2^a}} \sum_{j=1}^{n} \sum_{k=0}^{2^a-1} c_j\, e^{2\pi i k \lambda_j}\, |k\rangle |u_j\rangle .
\end{equation*}
Applying the inverse quantum Fourier transform to the first register maps each phase into a computational-basis state concentrated near $2^a\lambda_j$, so the state can be written as
$\sum_{j=1}^{n} c_j |\tilde{\lambda}_j\rangle |u_j\rangle$.
 
Because the components $|u_j\rangle$ are orthonormal, the distribution of the measured phase, conditioned on the component $|u_j\rangle$, depends only on $\lambda_j$, and this component occurs with probability $|c_j|^2$.
Fix $j$, and let $\beta_j$ be the integer in $\{0,1,\dots,2^a-1\}$ such that $\beta_j/2^a$ is the largest $a$-bit approximation to $\lambda_j$ not exceeding $\lambda_j$, so that $0 \le \lambda_j - \beta_j/2^a < 2^{-a}$. Let $m \in \{0,1,\dots,2^a-1\}$ denote the measurement outcome.
Following
\citet{nielsen2010quantum}, for any positive integer $e$,
\begin{equation*}
    \Pr\!\left(|m - \beta_j| > e \mid u_j\right) \le \frac{1}{2(e-1)} .
\end{equation*}
To approximate $\lambda_j$ to accuracy $2^{-b}$, we choose $e = 2^{a-b} - 1$, which gives
\begin{equation*}
    \Pr\!\left(|m - \beta_j| > e \mid u_j\right) \le \frac{1}{2(2^{a-b} - 2)} .
\end{equation*}
Writing $a = b + p$, where $p$ is the number of additional qubits used to control the failure probability, we obtain
\begin{equation*}
    \Pr\!\left(|m - \beta_j| \le e \mid u_j\right) \ge 1 - \frac{1}{2(2^{p} - 2)} .
\end{equation*}
Multiplying by the probability $|c_j|^2$ of the component $|u_j\rangle$, the joint probability of obtaining $|u_j\rangle$ together with a $b$-bit-accurate estimate of $\lambda_j$ is at least $|c_j|^2\left(1 - \frac{1}{2(2^p - 2)}\right)$. This is at least $|c_j|^2(1-\epsilon)$ whenever $\frac{1}{2(2^p-2)} \le \epsilon$, that is, whenever $p \ge \log_2\!\left(2 + \frac{1}{2\epsilon}\right)$. Since $p$ must be an integer, it suffices to choose $p = \left\lceil \log_2\!\left(2 + \frac{1}{2\epsilon}\right) \right\rceil$, so that $a = b + \left\lceil \log_2\!\left(2 + \frac{1}{2\epsilon}\right) \right\rceil$.
\end{proof}

To distinguish the eigenvalues without mixing them up, the measurement must be sufficiently sharp. The smallest distance between two eigenvalues determines how many precision bits $b$ are needed, which we calculate below.

\subsection{Precision Requirements and Qubit Allocation}
\label{sec:precision_requirements}
Before analyzing the execution of QPE over a complete orthonormal basis, we specify the precision $b$ required to resolve all distinct eigenvalues of the normalized Hamiltonian. Let $\lambda_1 < \lambda_2 < \dots < \lambda_d$ ($d \le n$) denote the distinct eigenvalues of the normalized Hamiltonian $H$, and define the minimum spectral gap
\begin{equation*}
    \Delta_{\min} = \min_{j \neq l} |\lambda_j - \lambda_l|.
\end{equation*}
Since $\operatorname{spec}(H) \subset (-0.5, 0.5)$, the $d-1$ consecutive gaps sum to $\lambda_d - \lambda_1 < 1$, so $0 < \Delta_{\min} < 1/(d-1)$ for $d \ge 2$.
 
A $b$-bit QPE estimate satisfies $|\tilde{\lambda}_j - \lambda_j| \le 2^{-b}$. To ensure that the estimation intervals of two adjacent distinct eigenvalues do not overlap, we require
\begin{equation*}
    2 \cdot 2^{-b} < \Delta_{\min}
    \quad\Longleftrightarrow\quad
    b > \log_2\!\left(\frac{2}{\Delta_{\min}}\right),
\end{equation*}
which is satisfied by
\begin{equation*}
    b = \left\lfloor \log_2\!\left(\frac{2}{\Delta_{\min}}\right) \right\rfloor + 1.
\end{equation*}
Choosing $b$ requires knowledge of $\Delta_{\min}$. We therefore assume that a lower bound $\Delta_{\mathrm{low}} \le \Delta_{\min}$ is known in advance, for example from the structure of $H$, as for the matrix families discussed in the Introduction. Using $\Delta_{\mathrm{low}}$ in place of $\Delta_{\min}$ only increases $b$, so the separation condition above continues to hold, and all complexity bounds below remain valid with $\Delta_{\min}$ replaced by $\Delta_{\mathrm{low}}$.

\subsection{Spectrum Estimation Using a Complete Orthonormal Basis}

We next show how the spectrum of $H$ can be obtained by executing the QPE procedure sequentially on the $n$ states of a predefined complete orthonormal basis. 
By using these basis states as inputs to the eigenvector register, we demonstrate that the resulting QPE measurements provide access to all eigenvalues and their corresponding eigenvectors without requiring any prior knowledge of the eigenvectors.

\begin{theorem}
Let $H$ be an $n$-dimensional sparse Hermitian matrix with orthonormal eigenvectors $\{\ket{u_j}\}_{j=1}^{n}$ and corresponding eigenphases $\{\lambda_j\}_{j=1}^{n}$. Let $\{\ket{v_i}\}_{i=1}^{n}$ be a predefined complete orthonormal basis used as input states for quantum phase estimation (QPE). If
\begin{equation*}
a=b+\left\lceil
\log_2\left(2+\frac{1}{2\epsilon}\right)
\right\rceil,
\end{equation*}
then, for each input state $\ket{v_i}$, the probability of obtaining a $b$-bit-accurate estimate of the eigenphase $\lambda_j$, together with the corresponding eigenstate component $\ket{u_j}$, is at least
\begin{equation*}
|\braket{u_j|v_i}|^2(1-\epsilon),
\end{equation*}
where $\epsilon\in(0,1)$ is the allowed failure probability in the phase-estimation accuracy guarantee.
Furthermore, since $\{\ket{v_i}\}_{i=1}^{n}$ forms a complete orthonormal basis, the completeness relation gives
\begin{equation*}
\sum_{i=1}^{n}|\braket{u_j|v_i}|^2=1.
\end{equation*}
Consequently, the sum of the lower bounds on the success probabilities over all input basis states satisfies
\begin{equation*}
\sum_{i=1}^{n}|\braket{u_j|v_i}|^2(1-\epsilon)
=1-\epsilon.
\end{equation*}
This establishes the accessibility of each eigenphase through the QPE executions. However, the sum of these individual lower bounds does not, by itself, equal the probability of detecting the eigenphase at least once.
\end{theorem}

\begin{proof}
Since $\{\ket{u_j}\}_{j=1}^{n}$ forms an orthonormal basis, each input state $\ket{v_i}$ can be expressed as
\begin{equation*}
\ket{v_i}=\sum_{j=1}^{n}c_j^i\ket{u_j},
\qquad
c_j^i=\braket{u_j|v_i}.
\end{equation*}
Consequently,
\begin{equation*}
|c_j^i|^2=|\braket{u_j|v_i}|^2.
\end{equation*}

For the $i$-th QPE execution, the input state is
\begin{equation*}
\ket{0}^{\otimes a}\otimes\ket{v_i}
=
\ket{0}^{\otimes a}\otimes
\sum_{j=1}^{n}c_j^i\ket{u_j}.
\end{equation*}
By the QPE procedure, the resulting state can be expressed, up to the phase-estimation error, as
\begin{equation*}
\sum_{j=1}^{n}c_j^i
\ket{\tilde{\lambda}_j}\ket{u_j},
\end{equation*}
where $\ket{\tilde{\lambda}_j}$ denotes the estimated eigenphase associated with $\ket{u_j}$.
By the assumed QPE accuracy guarantee, choosing
\begin{equation*}
a=b+\left\lceil
\log_2\left(2+\frac{1}{2\epsilon}\right)
\right\rceil
\end{equation*}
ensures that, for each $i$ and $j$, the probability of obtaining a $b$-bit-accurate estimate of $\lambda_j$, together with the corresponding eigenstate component $\ket{u_j}$, is at least
\begin{equation*}
|\braket{u_j|v_i}|^2(1-\epsilon).
\end{equation*}
We now show that every eigenstate has nonzero overlap with at least one input basis state. Since $\{\ket{v_i}\}_{i=1}^{n}$ is a complete orthonormal basis, it satisfies
\begin{equation*}
\sum_{i=1}^{n}\ket{v_i}\bra{v_i}=I.
\end{equation*}
Therefore, for any fixed $j$,
\begin{equation*}
\begin{aligned}
\sum_{i=1}^{n}|\braket{u_j|v_i}|^2
&=
\bra{u_j}
\left(\sum_{i=1}^{n}\ket{v_i}\bra{v_i}\right)
\ket{u_j}\\
&=\bra{u_j}I\ket{u_j}\\
&=1.
\end{aligned}
\end{equation*}

Multiplying both sides by $1-\epsilon$, we obtain
\begin{equation*}
\sum_{i=1}^{n}
|\braket{u_j|v_i}|^2(1-\epsilon)
=1-\epsilon.
\end{equation*}

Thus, every eigenstate $\ket{u_j}$ has a nonzero overlap with at least one input basis state, and its corresponding eigenphase $\lambda_j$ is accessible through the QPE executions with the prescribed phase-estimation accuracy. The sum above represents the sum of the individual lower bounds on the success probabilities; it does not, by itself, equal the probability of detecting the eigenphase at least once.

Since the argument holds for every $j\in\{1,\ldots,n\}$, every eigenphase is accessible through QPE without prior knowledge of the eigenvectors.
\end{proof}

To establish a high-probability guarantee for recovering the complete spectrum, we first examine the probability of detecting an individual eigenvalue when QPE is executed once for each input basis state. For this purpose, we consider the ideal phase-estimation setting by temporarily assuming that $\epsilon=0$. Let
\begin{equation*}
p_{ji}=|c_j^i|^2=|\langle u_j|v_i\rangle|^2
\end{equation*}
denote the probability of detecting the eigenvalue associated with $\ket{u_j}$ in the $i$-th QPE execution, assuming that the eigenphases are distinguishable. Under the assumption of independent executions, the probability of detecting this eigenvalue at least once across all $n$ executions is given by
\begin{equation*}
P_j=1-\prod_{i=1}^{n}(1-p_{ji}).
\end{equation*}
Although the completeness relation implies that
\begin{equation*}
\sum_{i=1}^{n}p_{ji}=1,
\end{equation*}
this identity alone does not guarantee that $P_j$ is arbitrarily close to one. To demonstrate this limitation, consider an eigenvector whose squared overlap with each computational basis state is uniform, i.e.,
\begin{equation*}
|\langle u_j|v_i\rangle|^2=\frac{1}{n},
\qquad i=1,\ldots,n.
\end{equation*}
In this case, each QPE execution detects the corresponding eigenvalue with probability $\frac{1}{n}$. Therefore, the probability of detecting the eigenvalue at least once after $n$ independent executions is
\begin{equation*}
P_j=1-\left(1-\frac{1}{n}\right)^n.
\end{equation*}
Taking the limit as $n\to\infty$ yields
\begin{equation*}
\lim_{n\to\infty}P_j=1-e^{-1}\approx 0.632.
\end{equation*}

This example demonstrates that, despite the squared overlaps summing to unity, the probability of detecting an individual eigenvalue can remain bounded away from one when QPE is executed only once for each input basis state. Consequently, the completeness relation alone does not guarantee recovery of the complete spectrum with arbitrarily high probability. This observation motivates the introduction of repeated QPE executions for each input basis state, enabling us to derive a stronger probabilistic guarantee for spectrum recovery.

\begin{theorem}
\label{theorem_4}
Let $H$ be an $n$-dimensional sparse Hermitian matrix with orthonormal eigenvectors $\{\ket{u_j}\}_{j=1}^{n}$ and corresponding distinguishable eigenphases $\{\lambda_j\}_{j=1}^{n}$. Let $\{\ket{v_i}\}_{i=1}^{n}$ be a predefined complete orthonormal basis, and define
\begin{equation*}
p_{ji}=|\braket{u_j|v_i}|^2.
\end{equation*}
Suppose that QPE is independently executed $R$ times for each input state $\ket{v_i}$, and that each execution yields a valid estimate of $\psi_j$ with probability at least $p_{ji}(1-\epsilon)$, where $0\leq\epsilon<1$. Then the probability of detecting all $n$ eigenphases is at least
\begin{equation*}
1-ne^{-R(1-\epsilon)}.
\end{equation*}
Consequently, for any $0<\delta<1$, it is sufficient to choose
\begin{equation*}
R\geq
\left\lceil
\frac{\ln(n/\delta)}{1-\epsilon}
\right\rceil
\end{equation*}
to ensure that all eigenphases are detected with probability at least $1-\delta$.
\end{theorem}

\begin{proof}
Fix an eigenphase $\lambda_j$. By assumption, for each input state $\ket{v_i}$, the probability of obtaining a valid estimate of $\lambda_j$ in a single QPE execution is at least
\begin{equation*}
p_{ji}(1-\epsilon).
\end{equation*}
Therefore, the probability of failing to detect $\lambda_j$ in all $R$ independent executions associated with $\ket{v_i}$ is at most
\begin{equation*}
\left[1-p_{ji}(1-\epsilon)\right]^R.
\end{equation*}
Since QPE is executed $R$ times for each of the $n$ input basis states, the probability of missing $\lambda_j$ in all executions satisfies
\begin{equation*}
P(\mathrm{miss}\ \lambda_j)
\leq
\prod_{i=1}^{n}
\left[1-p_{ji}(1-\epsilon)\right]^R.
\end{equation*}
Using the inequality $1-x\leq e^{-x}$ for $x\geq0$, we obtain
\begin{equation*}
\begin{aligned}
P(\mathrm{miss}\ \lambda_j)
&\leq
\exp\left[-R(1-\epsilon)
\sum_{i=1}^{n}p_{ji}\right].
\end{aligned}
\end{equation*}
Because $\{\ket{v_i}\}_{i=1}^{n}$ is a complete orthonormal basis,
\begin{equation*}
\sum_{i=1}^{n}p_{ji}
=
\sum_{i=1}^{n}|\braket{u_j|v_i}|^2
=1.
\end{equation*}
Hence,
\begin{equation*}
P(\mathrm{miss}\ \lambda_j)
\leq e^{-R(1-\epsilon)}.
\end{equation*}

Let $E_j$ denote the event that the eigenphase $\lambda_j$ is not detected in any execution. By the union bound,
\begin{equation*}
\begin{aligned}
P(\mathrm{miss\ any\ eigenphase})
&=P\left(\bigcup_{j=1}^{n}E_j\right)\\
&\leq\sum_{j=1}^{n}P(E_j)\\
&\leq ne^{-R(1-\epsilon)}.
\end{aligned}
\end{equation*}
Therefore, the probability of detecting all eigenphases is at least
\begin{equation*}
P(\mathrm{detect\ all\ eigenphases})
\geq1-ne^{-R(1-\epsilon)}.
\end{equation*}

To ensure that this probability is at least $1-\delta$, it suffices to require
\begin{equation*}
ne^{-R(1-\epsilon)}\leq\delta.
\end{equation*}
Taking natural logarithms and rearranging gives
\begin{equation*}
R\geq\frac{\ln(n/\delta)}{1-\epsilon}.
\end{equation*}
Since $R$ must be an integer, the stated choice of $R$ suffices. This completes the proof.
\end{proof}

\subsection{Complexity Analysis}
\label{Complexity Analysis}
The classical preprocessing step normalizes the sparse Hermitian matrix $H$. Since $H$ is $s$-sparse, its norm (such as $\|H\|_1$ or $\|H\|_F$) can be computed in $\mathcal{O}(ns)$ classical operations. The normalized Hamiltonian is then used to construct the unitary operator $U_H = e^{2\pi i H}$.
Simulating the evolution operator $e^{i H t}$ for duration $t$ requires gate complexity~\cite{Berry_2006, Harrow_2009}
\begin{equation*}
\mathcal{O}\left( \log(n) (\log^* n)^2 s^2 t \cdot 9^{\sqrt{\log(s^2 t / \epsilon_H)}} \right) = \tilde{\mathcal{O}}\left( s^2 t \log n \right),
\end{equation*}
where $\epsilon_H$ is the Hamiltonian-simulation error.
Here,
\begin{equation*}
\log^*n
\equiv
\min\left\{
r:\log_2^{(r)}n<2
\right\},
\end{equation*}
where $\log_2^{(r)}$ denotes the $r$-fold iterated logarithm, and $\tilde{\mathcal{O}}(\cdot)$ suppresses polylogarithmic and other slowly growing factors.

In standard Quantum Phase Estimation (QPE), implementing the sequence of controlled operations $U_H^{2^k} = e^{2\pi i 2^k H}$ for $k = 0, \ldots, a-1$ requires Hamiltonian simulation over a total cumulative evolution time proportional to
\begin{equation*}
t_{\text{total}} = t_0 \sum_{k=0}^{a-1} 2^k = t_0 (2^a - 1) = \mathcal{O}\left( \frac{t_0}{\Delta_{\min} \cdot \epsilon} \right),
\end{equation*}
where $t_0 \ge 2\pi$ is the base evolution time, and $a = \lceil \log_2(\frac{2}{\Delta_{\min}}) \rceil + \lceil \log_2(2 + \frac{1}{2\epsilon}) \rceil$. For a fixed target error probability $\epsilon$ and constant base time $t_0$, the total simulation time per QPE run scales as $\mathcal{O}\left(\frac{1}{\Delta_{\min}}\right)$.
Consequently, the complexity of the Hamiltonian simulation step within a single QPE execution is
\begin{equation*}
\tilde{\mathcal{O}}\left( \frac{s^2 \log n}{\Delta_{\min}} \right).
\end{equation*}
The inverse Quantum Fourier Transform (QFT) on the $a$ ancilla qubits contributes an additional $\mathcal{O}(a^2) = \mathcal{O}\left(\log^2\left(\frac{1}{\Delta_{\min}}\right)\right)$ elementary gates~\cite{Nielsen_Chuang_2010}, which is strictly dominated by the simulation cost.

To guarantee that all distinct eigenvalues and their corresponding eigenstates are detected with probability at least $1 - \delta$, QPE is executed $R = \mathcal{O}(\log(\frac{n}{\delta}))$ times for each of the $n$ computational basis states. Thus, the total number of QPE runs across all basis states is
\begin{equation*}
nR = \mathcal{O}(n \log n).
\end{equation*}
Multiplying the single-run simulation cost by the total number of executions $nR$, and including the one-time classical normalization cost $\mathcal{O}(ns)$, the overall runtime of the algorithm is
\begin{equation*}
\tilde{\mathcal{O}}\left( \frac{n s^2 \log^2(n)}{\Delta_{\min}} \right).
\end{equation*}
For inverse minimum spectral gaps up to $\Delta_{\min}^{-1} = \mathcal{O}(n)$,
this becomes $\tilde{\mathcal{O}}(n^2 s^2 \log^2 n)$, which improves on the
$\mathcal{O}(n^3)$ classical cost of computing all eigenvalues and
eigenvectors for sparsity up to $s = o(\sqrt{n}/\log n)$.

\section{Conclusion and Future Directions}
\label{Conclusion and Future Directions}
In this work, we presented a quantum algorithm for determining the spectrum of a sparse Hermitian matrix, including its eigenvalues and corresponding eigenvectors.
For an $n\times n$, $s$-sparse Hermitian matrix, the proposed algorithm has a runtime of $\tilde{\mathcal{O}}\left(
\frac{ns^2\log^2(n)}{\Delta_{\min}}
\right)$,
where $\Delta_{\min}$ denotes the minimum spectral gap.
Thus, for inverse minimum spectral gaps up to $\Delta_{\min}^{-1} = \mathcal{O}(n)$, and under the stated sparsity and input-access assumptions, the proposed algorithm achieves an improved asymptotic runtime compared with the $\mathcal{O}(n^3)$ time required by conventional classical algorithms for computing all eigenvalues and eigenvectors of an $n \times n$ matrix.
This regime includes the Hamiltonians of spin chains engineered for perfect quantum state transfer~\cite{christandl2004perfect}, as well as spin-$j$ angular momentum operators and the closely related Sylvester-Kac matrix~\cite{kac1947random,clement1959class}, all of which have equally spaced spectra.

Several directions remain for future work. First, replacing the Hamiltonian-simulation subroutine with qubitization-based methods and more efficient phase-estimation schemes could reduce the circuit depth. 
Second, the requirement of a known lower bound on $\Delta_{\min}$ could be removed by adaptively increasing the precision, and a rigorous procedure for filtering spurious phase estimates would strengthen the recovery guarantee. 
Third, we plan to extend the framework to non-Hermitian matrices through efficient Hermitian embeddings. Finally, we intend to validate the algorithm numerically on structured matrices such as the Sylvester-Kac matrix, and to explore applications of spectrum estimation in quantum simulation, the solution of differential equations, and other problems in scientific computing.

\bibliographystyle{splncs04}
\bibliography{ref} 

\begin{thebibliography}{10}
\providecommand{\url}[1]{\texttt{#1}}
\providecommand{\urlprefix}{URL }
\providecommand{\doi}[1]{https://doi.org/#1}

\bibitem{abrams1999quantum}
Abrams, D.S., Lloyd, S.: Quantum algorithm providing exponential speed increase for finding eigenvalues and eigenvectors. Physical Review Letters  \textbf{83}(24),  5162--5165 (1999). \doi{10.1103/PhysRevLett.83.5162}

\bibitem{RevModPhys.90.015002}
Albash, T., Lidar, D.A.: Adiabatic quantum computation. Rev. Mod. Phys.  \textbf{90},  015002 (Jan 2018). \doi{10.1103/RevModPhys.90.015002}

\bibitem{alman2025more}
Alman, J., Duan, R., Williams, V.V., Xu, Y., Xu, Z., Zhou, R.: More Asymmetry Yields Faster Matrix Multiplication, pp. 2005--2039. \doi{10.1137/1.9781611978322.63}

\bibitem{Berry_2006}
Berry, D.W., Ahokas, G., Cleve, R., Sanders, B.C.: Efficient quantum algorithms for simulating sparse hamiltonians. Communications in Mathematical Physics  \textbf{270}(2),  359–371 (Dec 2006). \doi{10.1007/s00220-006-0150-x}

\bibitem{RevModPhys.94.015004}
Bharti, K., Cervera-Lierta, A., Kyaw, T.H., Haug, T., Alperin-Lea, S., Anand, A., Degroote, M., Heimonen, H., Kottmann, J.S., Menke, T., Mok, W.K., Sim, S., Kwek, L.C., Aspuru-Guzik, A.: Noisy intermediate-scale quantum algorithms. Rev. Mod. Phys.  \textbf{94},  015004 (Feb 2022). \doi{10.1103/RevModPhys.94.015004}

\bibitem{cerezo2021variational}
Cerezo, M., Arrasmith, A., Babbush, R., Benjamin, S.C., Endo, S., Fujii, K., McClean, J.R., Mitarai, K., Yuan, X., Cincio, L., et~al.: Variational quantum algorithms. Nature Reviews Physics  \textbf{3}(9),  625--644 (2021). \doi{10.1038/s42254-021-00348-9}

\bibitem{christandl2004perfect}
Christandl, M., Datta, N., Ekert, A., Landahl, A.J.: Perfect state transfer in quantum spin networks. Physical Review Letters  \textbf{92},  187902 (2004). \doi{10.1103/PhysRevLett.92.187902}

\bibitem{chung1997spectral}
Chung, F.: Spectral Graph Theory. Conference Board of Mathematical Sciences, American Mathematical Society (1997). \doi{10.1090/cbms/092}

\bibitem{clement1959class}
Clement, P.A.: A class of triple-diagonal matrices for test purposes. SIAM Review  \textbf{1}(1),  50--52 (1959). \doi{10.1137/1001006}

\bibitem{dhillon2006design}
Dhillon, I.S., Parlett, B.N., V{\"o}mel, C.: The design and implementation of the {MRRR} algorithm. ACM Transactions on Mathematical Software  \textbf{32}(4),  533--560 (2006). \doi{10.1145/1186785.1186788}

\bibitem{farhi2000quantumcomputationadiabaticevolution}
Farhi, E., Goldstone, J., Gutmann, S., Sipser, M.: Quantum computation by adiabatic evolution. arXiv: Quantum Physics  (2000). \doi{10.48550/arXiv.quant-ph/0001106}

\bibitem{PhysRevA.89.022342}
Gaitan, F., Clark, L.: Graph isomorphism and adiabatic quantum computing. Phys. Rev. A  \textbf{89},  022342 (Feb 2014). \doi{10.1103/PhysRevA.89.022342}

\bibitem{golub2013matrix}
Golub, G., Van~Loan, C.: Matrix Computations. Johns Hopkins Studies in the Mathematical Sciences, Johns Hopkins University Press (2013). \doi{10.56021/9781421407944}

\bibitem{grover1996fast}
Grover, L.K.: A fast quantum mechanical algorithm for database search. In: Proceedings of the twenty-eighth annual ACM symposium on Theory of computing. pp. 212--219 (1996). \doi{10.1145/237814.237866}

\bibitem{Harrow_2009}
Harrow, A.W., Hassidim, A., Lloyd, S.: Quantum algorithm for linear systems of equations. Phys. Rev. Lett.  \textbf{103},  150502 (Oct 2009). \doi{10.1103/PhysRevLett.103.150502}

\bibitem{Horn_Johnson_2012}
Horn, R.A., Johnson, C.R.: Hints for Problems, p. 579–606. Cambridge University Press (2012). \doi{10.1017/CBO9781139020411}

\bibitem{J__2022}
J., A., Adedoyin, A., Ambrosiano, J., Anisimov, P., Casper, W., Chennupati, G., Coffrin, C., Djidjev, H., Gunter, D., Karra, S., Lemons, N., Lin, S., Malyzhenkov, A., Mascarenas, D., Mniszewski, S., Nadiga, B., O’malley, D., Oyen, D., Pakin, S., Prasad, L., Roberts, R., Romero, P., Santhi, N., Sinitsyn, N., Swart, P.J., Wendelberger, J.G., Yoon, B., Zamora, R., Zhu, W., Eidenbenz, S., B\"{a}rtschi, A., Coles, P.J., Vuffray, M., Lokhov, A.Y.: Quantum algorithm implementations for beginners  \textbf{3}(4) (Jul 2022). \doi{10.1145/3517340}

\bibitem{kac1947random}
Kac, M.: Random walk and the theory of {B}rownian motion. The American Mathematical Monthly  \textbf{54}(7),  369--391 (1947). \doi{10.2307/2304386}

\bibitem{lanczos1950iteration}
Lanczos, C.: An iteration method for the solution of the eigenvalue problem of linear differential and integral operators. Journal of Research of the National Bureau of Standards  \textbf{45},  255--282 (1950). \doi{10.6028/jres.045.026}

\bibitem{mermin2007quantum}
Mermin, N.D.: Quantum computer science: an introduction. Cambridge University Press (2007). \doi{10.1017/CBO9780511813870}

\bibitem{horowitz2019quantum}
{National Academies of Sciences, Engineering, and Medicine}: Quantum Computing: Progress and Prospects. The National Academies Press, Washington, DC (2019). \doi{10.17226/25196}

\bibitem{Nielsen_Chuang_2010}
Nielsen, M., Chuang, I.: Quantum Computation and Quantum Information: 10th Anniversary Edition. Cambridge University Press (2010). \doi{10.1017/CBO9780511976667}

\bibitem{nielsen2010quantum}
Nielsen, M.A., Chuang, I.L.: Quantum computation and quantum information. Cambridge university press (2010). \doi{10.1080/00107514.2011.587535}

\bibitem{paige1980accuracy}
Paige, C.C.: Accuracy and effectiveness of the {L}anczos algorithm for the symmetric eigenproblem. Linear Algebra and its Applications  \textbf{34},  235--258 (1980). \doi{10.1016/0024-3795(80)90167-6}

\bibitem{parlett1998symmetric}
Parlett, B.N.: The Symmetric Eigenvalue Problem. Classics in Applied Mathematics, SIAM (1998). \doi{10.1137/1.9781611971163}

\bibitem{pastorello2022concise}
Pastorello, D.: Concise Guide to Quantum Machine Learning. Machine Learning: Foundations, Methodologies, and Applications, Springer Nature Singapore (2022). \doi{10.1007/978-981-19-6897-6}

\bibitem{peruzzo2014variational}
Peruzzo, A., McClean, J., Shadbolt, P., Yung, M.H., Zhou, X.Q., Love, P.J., Aspuru-Guzik, A., O’brien, J.L.: A variational eigenvalue solver on a photonic quantum processor. Nature communications  \textbf{5}(1), ~4213 (2014). \doi{10.1038/ncomms5213}

\bibitem{Preskill2018quantumcomputingin}
Preskill, J.: Quantum {C}omputing in the {NISQ} era and beyond. {Quantum}  \textbf{2}, ~79 (Aug 2018). \doi{10.22331/q-2018-08-06-79}

\bibitem{saad2011numerical}
Saad, Y.: Numerical Methods for Large Eigenvalue Problems. Classics in Applied Mathematics, SIAM, revised edn. (2011). \doi{10.1137/1.9781611970739}

\bibitem{shor1994algorithms}
Shor, P.W.: Algorithms for quantum computation: discrete logarithms and factoring. In: Proceedings 35th annual symposium on foundations of computer science. pp. 124--134. Ieee (1994). \doi{10.1109/SFCS.1994.365700}

\bibitem{trefethen1997numerical}
Trefethen, L.N., Bau, III, D.: Numerical linear algebra. SIAM (1997). \doi{10.1137/1.9780898719574}

\bibitem{10.1145/1536414.1536452}
Trevisan, L.: Max cut and the smallest eigenvalue. In: Proceedings of the Forty-First Annual ACM Symposium on Theory of Computing. p. 263–272. STOC '09, Association for Computing Machinery, New York, NY, USA (2009). \doi{10.1145/1536414.1536452}

\bibitem{Zaman_2023}
Zaman, A., Morrell, H.J., Wong, H.Y.: A step-by-step hhl algorithm walkthrough to enhance understanding of critical quantum computing concepts. IEEE Access  \textbf{11},  77117–77131 (2023). \doi{10.1109/access.2023.3297658}

\end{thebibliography}
\end{document}